\documentstyle[aps,preprint,tighten,floats,epsf,rotate]{revtex}

\begin{document}
\draft


\title{Measuring Cosmic Defect Correlations in Liquid Crystals}

\author{Rajarshi Ray \footnote{email: rajarshi@iopb.res.in} and Ajit M. 
Srivastava \footnote{email: ajit@iopb.res.in, present address (until
January 2002): Physics Dept., University of California, Santa Barbara,
CA, USA}}

\address{Institute of Physics, Sachivalaya Marg, Bhubaneswar 751005, 
India}


\maketitle
\widetext

\parshape=1 0.75in 5.5in
\begin{abstract}
From the theory of topological defect formation proposed for the early 
universe, the so  called Kibble mechanism, it follows that the density
correlation functions of defects and anti-defects in a given system should 
be completely determined in terms of a single length scale $\xi$, the 
relevant domain size, which is proportional to the average inter-defect
separation $r_{av}$. Thus, when lengths are expressed in units of $r_{av}$, 
these distributions should show universal behavior, depending only on the 
symmetry of the order parameter, and space dimensions. We have verified 
this prediction by analyzing the distributions of defects/anti-defects
formed during the isotropic-nematic phase  transition in a thin layer in 
a liquid crystal sample. Our experimental results confirm this prediction
and are in reasonable agreement with the results of  numerical simulations.
\end{abstract}
\vskip 0.125 in
\parshape=1 -.75in 5.5in
\pacs{PACS numbers: 61.30.Jf, 98.80.Cq, 64.70.Md}
\narrowtext
 
 The Big-Bang theory of the universe is well established by now,
with most of the observations being in good agreement with the
predictions of the model. Accurate measurements of the cosmic
microwave background radiation (CMBR) have made cosmology into
a precision science where competing models of the early universe
are put to rigorous tests. One of the models which was very popular
earlier for explaining the formation of structure in the universe,
such as galaxies, clusters, and superclusters of galaxies, utilized
the concept of the {\it topological defects}, in particular cosmic 
string defects \cite{csbook}. Though recent CMBR anisotropy data is 
not in agreement with the predictions of topological defects based 
models of structure formation, it will be premature to 
completely reject these models due to various unresolved issues 
regarding these models \cite{tdcmbr}. Further, topological defects arise
in many particle physics models of unification of forces, and their
presence in the early universe can lead to other important consequences,
such as, production of baryon number below electroweak scale, generating
baryon number inhomogeneities at quark-hadron transition, etc. 
\cite{dfluc}. It is therefore important to deepen our understanding of 
how these defects form in the universe \cite{tdform}, and how they evolve. 
This paper relates to first of these issues. 

 There are numerous examples of topological defects in condensed matter 
systems, such as, flux tubes in superconductors, vortices in superfluid 
helium, monopoles and strings in liquid crystals, etc. It was long known 
that such defects routinely form during a phase transition. First detailed 
theory of formation of topological defects in a phase transition (apart
from the usual equilibrium process of thermal production) was proposed by 
Kibble \cite{kbl,kbl2} in the context of the early Universe. It is usually 
referred to as {\it the Kibble mechanism}. It was first suggested by Zurek 
\cite{zrk1} that some of the aspects of the Kibble mechanism can be tested 
in condensed matter systems, such as superfluid $^4$He. Indeed the basic 
physical picture of the Kibble mechanism applies equally well to a 
condensed matter system  \cite{zrk2,rjnt,tdform}. 
Using this correspondence, the basic picture of the defect formation
was first observed by Chuang et al. in isotropic to nematic (I-N) 
transition in liquid crystal systems \cite{lc1}.
Formation and evolution of string defects was observed to be remarkably
similar to what had been predicted for the case of the universe, apart
from obvious differences such as the velocity of defects, time scales
of the evolution of defect network etc. Subsequently, measurements
of string density were carried out by observing strings formed in the
first order I-N transition occurring via the nucleation
of nematic bubbles in the isotropic background \cite{lc2}. The results
were found to be in good agreement with the prediction of the the
Kibble mechanism. There have been other experiments, with superfluid helium
\cite{he}, and with superconductors \cite{sccor}, where predictions of 
Kibble  mechanism have been put to test. Yet another quantitative measurement 
was made of the exponent charactering the correlation between the defects and
the anti-defects formed in the I-N transition in liquid crystals, with 
results in good agreement with the Kibble mechanism \cite{lc3}.

  It is certainly dramatic that there is a correspondence between the 
phenomenon expected to have occurred in the early universe, during stages 
when its temperature was about $10^{29}$ K, with those occurring in the 
condensed matter systems at temperatures less than few hundred K. 
The important question is whether the observations/measurements performed in 
condensed matter systems provide rigorous tests of the theories being used
to predict phenomenon in the early universe, or they simply provide
an analogy with the case of the early universe, or at best, examples of
other systems where {\it similar theoretical considerations} can be made
for investigating  defect formation. On the face of it, the
differences between the two cases seem so profound that a rigorous test 
of theories underlying the phenomenon taking place in the early universe
seems out of question within the domain of condensed matter systems. 
For example, the description of the matter, transformations between
its various phases etc. in the early universe is given in terms of 
elementary particle physics models, which require the framework of
relativistic quantum field theory. Indeed, typical velocities of particle,
and even the topological defects formed in such transitions, are 
close to the velocity of light. On the other hand, in condensed matter
systems non-relativistic quantum mechanics provides adequate description
of most relevant systems. Typical velocities in such systems are 
extremely small in comparison. For example, in liquid crystals, typical 
velocity of string defects is of the order of few microns per second, 
completely negligible in comparison to the speed of light. Similarly, 
the value of string  tension for the case of the universe is of the 
order of $10^{32}$ GeV$^2$ which is about $10^{18}$ tons/cm. In contrast, 
in condensed matter systems (say, in liquid crystals), the string 
tension is governed by the scale of the free energy of the order of 
$10^{-2}$ eV. Indeed, it is due to this extremely large string tension 
(i.e. mass per unit length) of cosmic string defects, that they were 
proposed to have seeded formation of galaxies, clusters of galaxies etc. 
via their gravitational effects. Clearly such effects are unthinkable 
for string defects in liquid crystals, or in superfluid helium etc.

  Despite the fact that the two systems look completely dissimilar, it
turns out, that there are ways in which specific condensed matter
experiments can provide rigorous tests of the theories of cosmic
defect formation. This can be done by identifying those predictions
of the theory which show universal behavior. As we will discuss below 
there are several predictions of the Kibble mechanism which show 
universal behavior when expressed in terms of suitable length scale. 

 We first describe the basic physics of the Kibble 
mechanism \cite{kbl,kbl2,rjnt}.
For concreteness, we consider the case of a complex scalar order parameter
$\phi$, with spontaneously broken U(1) symmetry (as in the case of 
superconductors, or superfluid $^4$He). The order parameter space (the
vacuum manifold) is a circle $S^1$ in this case. In the Kibble mechanism, 
defects form due to a domain structure arising in a phase transition. This 
domain like structure arises from the fact that during phase transition, 
the phase $\theta$ of the order parameter field $\phi$ can only be 
correlated within a finite region. $\theta$ can be taken to be  roughly 
uniform within a region (domain) of size $\sim \xi$,
while varying randomly from one domain to the other. This situation is
very natural to expect in a first order phase transition where the transition
to the spontaneous symmetry broken phase happens via nucleation of bubbles.
Inside a bubble, $\theta$ will be uniform, while $\theta$ will vary
randomly from one bubble to another. Eventually bubbles grow and coalesce,
leaving a region of space where $\theta$ varies randomly at a distance scale
of the inter-bubble separation, thereby leading to a domain like structure.
[In certain situations, e.g. when bubble wall motion is highly dissipative,
effective domain size may be larger \cite{dspt,tdform}.
In between any two adjacent bubbles (domains), $\theta$ is supposed to 
vary with least gradient. This is  usually called the $geodesic~rule$ and 
arises naturally from the  consideration of minimizing the gradient energy
for the case of global symmetry. For gauge symmetry, situation is more 
complicated \cite{gauge1}. Geodesic rule may not hold in the presence of 
strong fluctuations \cite{flip,gauge11,gauge2}. Recently it has been shown 
\cite{gauge2}, that a defect distribution very different from what is
expected in the Kibble mechanism, may arise when defect formation is dominated
by magnetic field fluctuations.]

The same situation happens for a second order transition where the orientation 
of the order parameter field is correlated only within a region of the size 
of the correlation length $\xi$. This again results in a domain like
structure, with domains being the correlation volumes. We mention here that
there are non-trivial issues in the case of a second order phase transition,
in determining the appropriate correlation length for calculating the 
{\it absolute initial defect density}. It was first pointed out by Kibble 
\cite{kbl2} (see also, \cite{tdform}) that the appropriate value of $\xi$ 
in this case should depend on the rate of phase transition. It was
discussed by Zurek \cite{zrk2} that the appropriate value of $\xi$ should
be determined by incorporating the effects of critical 
slowing down of the dynamics of the order parameter field near the transition 
temperature. The theory which takes 
into account of this (for second order transitions) is usually referred to as 
the {\it Kibble-Zurek} mechanism. Here, the absolute defect density is 
determined by critical fluctuations of the order parameter, and depends on 
details such as the rate of cooling of the system etc. For a discussion of 
these issues, see ref.\cite{zrk2}. We emphasize that these considerations of 
the details of the critical dynamics of the order parameter field 
during the phase transition are important in calculating the absolute defect 
density. However, defect density {\it per domain} is insensitive to 
these details.

\begin{figure}[h]
\begin{center}
\vskip -2in
\leavevmode
\epsfysize=15truecm \vbox{\epsfbox{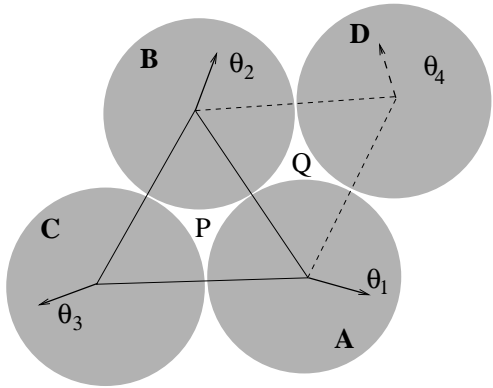}}
\end{center}
\vskip -2in
\caption{Defect formation due to coalescence of three domains.}
\label{Fig.1} 
\end{figure}

 For the U(1) case which we are discussing, string defects (vortices)
arise at the junctions of domains if $\theta$ winds non trivially around a
closed path going through adjacent domains. Consider a junction P of three
domains, as shown in Fig.1. For simplicity, we show here domains as
spheres (e.g. bubbles for a first order transition), with the centers
of the three nearest bubbles forming an equilateral triangle. The values 
of $\theta$ in these three domains are, $\theta_1$,
$\theta_2$ and $\theta_3$. One can show that with the use of the geodesic
rule (i.e., the variation of $\theta$ in between any two domains is along
the shortest path on the order parameter space $S^1$), a non-trivial
winding of $\theta$ (i.e. by $\pm 2\pi$) along a closed path,  encircling 
the junction P and going through the three domains A,B,C, will arise only 
when $\theta_3$
lies in the (shorter) arc between $\theta_1 + \pi$ and $\theta_2 + \pi$.
Maximum and minimum values of the angular span of this arc are
$\pi$ and 0, with average angular span being $\pi/2$. Since $\theta_3$
can lie anywhere in the circle, the probability that it lies in the
required range is $p = (\pi/2)/(2\pi) = 1/4$. Thus, we conclude \cite{tv}
that the probability of vortex (or antivortex) formation at a junction of
three domains (in 2-space dimensions) is equal to 1/4.

 It is important to realize that in the above argument, no use is
made of the field equations. Thus, whether the system is a relativistic
one for particle physics, or a non-relativistic one appropriate for 
condensed matter physics, there is no difference in the defect production 
per domain. [This is so as long as the geodesic rule holds. As we have
mentioned above, this may not be true when the field dynamics is dominated 
by fluctuations \cite{flip,gauge11,gauge2}. In those situations, defect 
production will be determined by some different processes, e.g. by a new 
{\it flipping mechanism} \cite{flip,gauge11}, or by fluctuating magnetic fields
\cite{gauge2}, again, apart from the usual equilibrium thermal production.] 
Therefore, the expected number of defects {\it per domain} has universal 
behavior, in the sense that it only depends on the symmetry of the order 
parameter and the space dimensions. It is this universal nature of the 
prediction of defect density (number of defects per domain) in the Kibble 
mechanism which has been utilized to experimentally verify this prediction 
(which was originally given for cosmic defect production) in liquid crystal 
systems \cite{lc2}. Note that absolute defect density does not show universal
behavior since it depends on the domain size $\xi$. The entire dependence
on the dynamical details of the specific system is through the single
length scale $\xi$. Thus, when densities are expressed in length scale
of $\xi$ the prediction acquires a universal character. 

  Kibble mechanism does not only predict the number density of defects. 
It also predicts a very specific correlation between the defects and 
anti-defects. Again, by focusing on specific quantities and choosing
proper length scales, this prediction acquires a universal nature which 
can then be tested experimentally in condensed matter systems. This prediction
is important as in many experimental situations it can provide the only
rigorous test of the underlying theory of defect formation. This is for the 
following reason. As we have discussed above, prediction of defect density 
becomes universal only when expressed in the length units of the domain size. 
For a first order transition it can be done easily, if the average bubble 
diameter (assuming small variance in bubble sizes) can be experimentally
determined, as in liquid crystal experiments in ref. \cite{lc2}). However,
this is a highly non-trivial problem for the case of second order
transition (or for spinodal decomposition). For experimental
situations with second order transition (type II superconductors, 
superfluid $^4$He), it has not been possible to make independent 
determination of the correlation length relevant for defect formation. All 
one can do is to measure absolute defect density, comparison of which to 
the  prediction from theory becomes dependent on various details of
phase transition, which determine the relevant correlation length
\cite{zrk2}.

  Defect-anti-defect correlations, on the other hand, imply specific
spatial distributions of defects, which are independent of the prediction
of defect density. Within the framework of the Kibble mechanism, defect
density distributions are expected to reflect defect-anti-defect correlations
at the typical distance scales of a domain size. At the same time, typical
inter-defect separation $r_{av}$, as given by $\rho^{-1/2}$, where $\rho$ 
is the average defect/anti-defect density, is directly proportional to the 
domain size $\xi$. From this, one can conclude that if defect and 
anti-defect distributions are expressed in terms of the average 
inter-defect separation $r_{av}$, then these should display universal 
behavior. The important point being that the same experiment yields the 
value of average inter-defect separation ($r_{av} = \rho^{-1/2}$), and 
the defect and anti-defect distributions are analyzed by using this
length scale. Details such as the bubble size, or the relevant 
correlation length, therefore, become completely immaterial for testing
the predictions regarding correlations. 

To understand how this correlation between defects and anti-defects arises, 
let us go back to Fig.1. With the directions of arrows shown there (which 
denote values of $\theta$), we see that a defect (vortex with winding
$+ 1$) has formed at P. Let us address the issue that, given that a defect 
has formed at P, how does the probability of a defect, or anti-defect, 
change in the nearest triangular region, say at Q. At Q, the three domains 
which intersect are A,B, and D. $\theta$ already has specific variation 
in A and B in order to yield a defect at P. From the point of view of 
Q, this variation in A and B is a partial winding configuration for an 
anti-vortex. With partial anti-winding present in A and B, it can be seen 
from Fig.1 that whatever be the value of $\theta$ in D, it is impossible 
to have a vortex at Q (i.e. $\theta$ winding by $+ 2\pi$ as we go around Q
in an anti-clockwise manner). On the other hand, the probability of
an antivortex formation at Q is 1/4 by straightforward repetition of
the argument given above to calculate the probability of formation of
a vortex {\it or} antivortex at a junction of three domains. This is
the most dramatic example of correlation between  defect and anti-defect
formation. If we consider defect formation due to intersection of 
(say) four domains, this correlation still exists, that is close to a 
defect, formation of another defect is less likely (though not completely
prohibited now) and formation of anti-defect is enhanced. This conclusion, 
about certain correlation in the formation of a defect and an anti-defect, 
is valid for other types of defects as well \cite{skrm}. 

 To see what this correlation implies, let us consider a two
dimensional region $\Omega$ whose area is $A$ and  whose perimeter 
{\it L} goes through $L/\xi$ number of elementary domains (where $\xi$ is 
the domain size). As $\theta$ varies randomly from one domain to another, 
one is essentially dealing with a random walk problem with the average 
step size for $\theta$ being $\pi/2$ (the largest step is $\pi$ and the
smallest is zero). Thus, the net winding number of $\theta$ around {\it L} 
will be distributed about zero with a typical width given by $\sigma = 
{1 \over 4} \sqrt{L \over \xi}$, implying that $\sigma \propto A^{1/4}$ 
(see, ref. \cite{vv,lc3,rjnt}). Assuming roughly uniform defect density, 
we get $\sigma \propto N^{1/4}$ (where $N$ is the total number of defects in 
the region $\Omega$), which reflects the correlation in the production of 
defects and anti-defects. In the absence of any correlations, the 
net defect number will not be as suppressed, and will follow 
Poisson distribution with $\sigma \sim \sqrt{N}$. In general
one may write the following scaling relation for $\sigma$,

\begin{equation}
\sigma ~ = ~ C ~ N^\nu
\end{equation}

 The exponent $\nu$ will be 1/2 for the uncorrelated case and 1/4 for the
case of the Kibble mechanism. Again, the prediction of $\nu = 1/4$ is
of universal nature, depending only on the symmetry of order parameter and
space dimensions. An experimental
measurement of this exponent was carried out in ref.\cite{lc3}. The 
experimental value of $\nu$ was found in ref.\cite{lc3} to be $\nu = 
0.26 \pm 0.11$ which is in a
good agreement with the predicted value of 1/4 from the Kibble 
mechanism, and reflects the correlated nature of defects and anti-defects. 

 The exponent $\nu$ does not give complete information about the
correlation which arises between defects and anti-defects when they are
produced via the Kibble mechanism. A more detailed understanding of
the correlation can be achieved by calculating the density
correlation function of the defects and anti-defects. Below, we
first discuss theoretical prediction about this, 
and then describe the experimental measurements. 

\begin{figure}[h]
\begin{center}
\vskip -2in
\leavevmode
\epsfysize=15truecm \vbox{\epsfbox{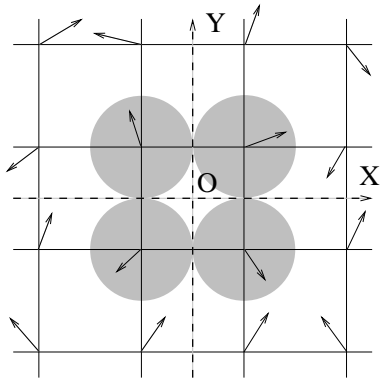}}
\end{center}
\vskip -2in
\caption{Defect formation due to coalescence of four elementary
domains, forming a square lattice.}
\label{Fig.2} 
\end{figure}

  We have carried out the numerical simulation of defect formation via the 
Kibble mechanism for the square lattice case (instead of triangular case as 
shown in Fig.1) where defects will form at the intersection of four domains, 
see Fig.2. This is because we find slightly better agreement with the 
experimental results for the square lattice case. We will also quote results 
for the triangular lattice case. The probability $p$ of a defect or an 
anti-defect per square region (in Fig.2) is obtained to be 0.33. We start with 
a defect at the origin O of the coordinate system. We then calculate the 
density of anti-defects $\rho_{\bar d}(r)$, as well as the density of defects 
$\rho_d(r)$, as a function of the radial distance $r$ from the defect 
at the origin. For this, we count number of anti-defects (defects) in an
annular region of width $\Delta r$ centered at distance $r$ from the origin, 
and divide this number by the area of the annular region. An average of these 
densities (at a given $r$) is taken by taking many realizations of the 
event of defect production. With density in the central square (which has a 
defect) appropriately normalized, this is the same as the density 
correlation function. As should be clear by  now, all the dynamical 
details of the specific model are relevant only in determining the domain 
size $\xi$ (i.e., either the bubble diameter for a first order transition 
case, or the relevant correlation length for a second order transition 
case, or for spinodal decomposition). For the simulation, we take
domain size to be unity, meaning that all lengths are measured in 
units of domain size. All remaining properties of the defect distributions
should now display universal behavior, depending only on the symmetry of the 
order parameter (U(1) in this case), space dimensions (2 here), and, possibly, 
fundamental domain structure (square vs. triangular). Another important factor 
here is that there is no reason to expect that the defect/anti-defect will be 
exactly at the center of the square formed by the centers of 
the four bubbles (domains). In realistic situation, the defects can be 
anywhere in the elementary square, which in some sense represents the 
collision region of the four correlation domains (bubbles). However, 
since the correlation domains, by definition, have uniform
order parameter, it should be increasingly unlikely that the defect is 
far from the center of the collision region (i.e. away from O). To take into
account of these physical considerations, we have allowed  the position of 
defects/anti-defects to float within a square with an approximately 
Gaussian probability distribution, centered in the middle of the square,
with varying width. By changing the width of this distribution we can 
change from almost centered defects, to defects with uniform probability 
in the elementary square region. 

  With this picture, and with a defect being in the central square, we know 
that the probability of anti-defect formation should be enhanced in the 
nearest squares, while the probability of defect formation should be 
suppressed in these regions. These defects/anti-defects cannot be too close
to the central defect as that would imply variation of the order parameter
at distance scales much smaller than $\xi$. One therefore expects that both 
$\rho_{\bar d}(r)$ and $\rho_d(r)$ will be almost zero for $r << \xi$
(discounting the central defect at O). Both densities will rise as $r$
increases. At a distance of about $r \sim \xi$, one expects a peak in
$\rho_{\bar d}(r)$. The height of this peak will determine the suppression
in $\rho_d(r)$ (compared to the asymptotic value) at that point. 

 For $r \ge 2 \xi$ one expects both distributions to approach the average
density expected asymptotically. This conclusion may appear surprising as
one might have expected that increased anti-defect probability in the nearest 
square may imply suppressed anti-defect probability in the next-nearest
square etc., leading to a damped oscillatory behavior of $\rho_{\bar d}(r)$
and $\rho_d(r)$ (similar to density correlation function for
a liquid). The reason that this does not happen here is simple, and again,
intrinsic to the Kibble mechanism. Recall that the anti-defects were enhanced
(and defects suppressed) in the squares nearest to the central square 
because for each of these four squares, two out of four vertices had 
$\theta$ common to the central square which already had a defect. The 
probability of defects/anti-defects in these squares was, therefore, 
affected by the presence of defect in the central square. In contrast, 
$\theta$ at three vertices of the {\it next nearest} (corner) squares, and at 
all vertices further away, are completely random. This implies that the 
probability of defect and anti-defect must be equal in all such regions, 
leading to flattening of $\rho_{\bar d}(r)$ and $\rho_d(r)$ for $r \ge 2\xi$.
As we will see below, this theoretical reasoning is well born out by 
the results of the simulations, and is consistent with the experimental
results, within error bars.

  We have explained above that this analysis of defect-anti-defect
correlations does not require the knowledge of domain size. In fact,
the power of this technique is best illustrated in those experimental
situations where domains are not identifiable (as in the experiments
explained below). The length scale $\xi$ is, therefore, not a convenient
choice from this point of view. We, instead, will use the inter-defect
separation $r_{av}$ to define our length scale. If the probability of
defect formation per (square) domain is $p$, then $r_{av}$ and $\xi$
are related in the following manner.

\begin{equation}
r_{av} = {\xi \over \sqrt{p}}
\end{equation}

In our experiment, we have tried to record the defect distributions as
soon as defects form. Note that even if some evolution of defect 
network occurs by coarsening of domains, Eq.2 still holds. Coarsening 
of domains only makes the effective correlation domain size $\xi$ larger.
Since the probability $p$ (defects per domain) is also of universal nature, 
it follows from Eq.(2) that the distributions $\rho_{\bar d}(r)$ and 
$\rho_d(r)$ will still show universal behavior if $r$ is measured in the
units of $r_{av}$. As $r_{av}$ can be directly measured for a given
defect distribution, without any recourse to underlying domain
structure, we will use it as defining the unit of length. In this
unit, the peak in $\rho_{\bar d}(r)$ will be expected to occur
at $r = \sqrt{p} \simeq 0.57$ with $p = 0.33$ for the square lattice case.
(This is when the order parameter space is $S^1$, which, as we will 
discuss below, is the case for our experiment). By $ r = 2 \sqrt{p}
\simeq 1.14$, both densities will be expected to flatten out to the 
asymptotic value of 0.5. The asymptotic value of $\rho_d$ and
$\rho_{\bar d}$ is 0.5 since by definition the unit length $r_{av}$
is the average separation between defects/anti-defects. The height of the
peak for $\rho_{\bar d}$ (at $r \simeq 0.57$) depends on the weight factor
for centering the defects/anti-defects inside square regions. This
is simply because perfectly centered defects/anti-defects will lead to
large contributions when centers of the squares fall inside the annular
regions for calculating densities. While for neighboring values of $r$,
their contributions will be zero. When defects/anti-defects positions are
floated, then the contributions are averaged out. We find that the height
of the peak varies from about 1.2 for perfectly centered defects/anti-defects 
to about 0.75 for the case when defects/anti-defects positions are uniformly 
distributed inside a square. When comparing with the experimental results,
we choose the weight factor appropriately so that the height of the
peak in $\rho_{\bar d}$ from simulation is similar to the one obtained from
experiment. (It will be interesting to investigate the physics contained
in the peak height resulting from this weight factor for centering
the defects in a square, which may depend on the interactions between
defects/anti-defects.) 

 If we take the thickness of the annular region $\Delta r$ (for calculating 
densities) to be about the domain size (i.e., 0.57 in units of $r_{av}$), 
then one can relate the suppression in $\rho_d$ at $r = \sqrt{p} \simeq 
0.57$ to the height of the peak in $\rho_{\bar d}$ at that position. This 
is done as follows. Note that the values of $\theta$ at all the outer 12 
vertices of the 8 squares, bordering the central square, are completely 
random (i.e. they are not constrained by the fact that there is a defect 
in the central square). Let us consider the distribution of the net 
winding along the large square shaped path going through all these 12 
vertices. Repetition of the earlier argument (in writing Eq.(1)) shows that
this net winding should be distributed about the value zero, (and should
have a typical width proportional to $\sqrt{12}$, but this part is
not relevant here). Thus, average number of defects $n_d$ inside this
large square should be same as the average number of anti-defects 
$n_{\bar d}$. However, we are given that the central square always has
one defect. With the area of each domain $\xi^2 = p$ in the units 
of $r_{av}$, this directly leads to the relation between $\rho_d$ and
$\rho_{\bar d}$ at $r \simeq 0.57$ to be

\begin{equation}
\rho_d(r = 0.57) = \rho_{\bar d}(r = 0.57) - {1 \over 8p} ~ \simeq ~
\rho_{\bar d}(r = 0.57) - 0.39,
\end{equation}

\noindent for $p = 0.33$. Here, we have approximated the densities at 
$r \simeq 0.57$ by dividing the expected number of defects (anti-defects)
in the eight squares bordering the central square, by the total area of 
these squares, even though the corner squares are somewhat further away.
(If one accounts for the larger distance of these corner squares, and that
defects and anti-defects are equally probably there, then the suppression
in $\rho_d(r = 0.57)$ would be larger.)

 Our experiments have been carried out using nematic liquid 
crystals. For uniaxial nematic liquid crystals (NLC) the orientation 
of the order parameter in the nematic phase is given by a unit vector 
(with identical opposite directions) called the director. The 
order parameter space is $RP^2 (\equiv S^2/Z_2)$, which allows
for string defects with strength 1/2 windings.
Due to birefringence of NLC, when the liquid crystal sample is placed 
between crossed polarizers, then regions where the director is either 
parallel, or perpendicular, to the electric field ${\vec E}$, the 
polarization is maintained resulting in a dark brush. At other regions, 
the polarization changes through the sample, resulting in a bright 
region. This implies that for a defect of strength $s$, one will observe 
$4s$ dark brushes \cite{chndr}. If the cross-polarizer setup 
is rotated then brushes will rotate in the same (opposite) direction for 
positive (negative) windings. Equivalently, if the sample is rotated 
between fixed crossed polarizers, then brushes do not rotate for +1 
winding while they rotate in the same direction (with twice the angle 
of rotation of the sample) for $-$1 winding. We have used this method 
to determine the windings.  

  We now describe our experiment.  We observed isotropic-nematic (I-N) 
transition in a tiny droplet (size $\sim$ 2-3 mm) of NLC 
$4'$-Pentyl-4-biphenyl-carbonitrile (98\% pure, purchased from Aldrich 
Chem.). The sample was placed on a clean, untreated glass slide and was 
heated using an ordinary lamp. The I-N transition temperature is about 
35.3 $^0$C.  Our setup allowed the possibility of slow heating, and 
cooling, by changing the distance of the lamp from the sample. We 
observed the defect production very close to the transition temperature
(in some cases we had some isotropic bubbles co-existing with the nematic 
layer containing defects). For the observations, we used a Leica, DMRM 
microscope with 20x objective, a CCD camera, and a cross-polarizer setup, 
at the Institute of Physics, Bhubaneswar.
Phase transition process was recorded on a standard video cassette 
recorder.  The images were photographed directly from a television 
monitor by replaying the cassette.

\begin{figure}[h]
\begin{center}
\leavevmode
\epsfysize=6.5truecm \vbox{\epsfbox{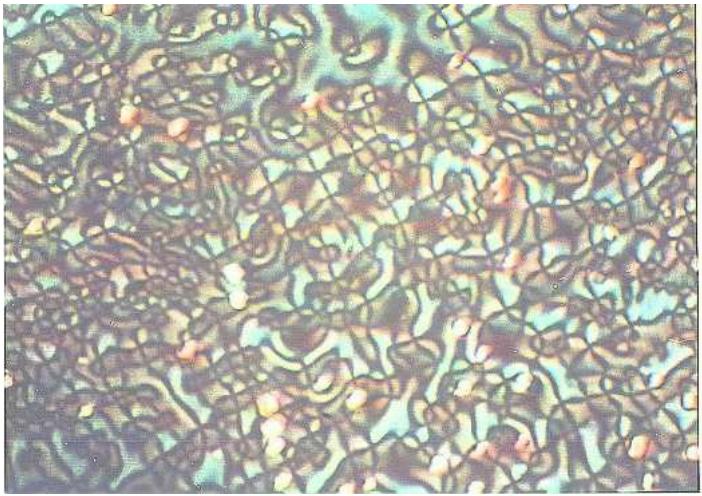}}
\end{center}
\caption{Network of strength one defects and anti-defects formed
in I-N transition. Crossing of brushes denotes defects with winding
$\pm$ 1. Size of the image is about 0.45 mm $\times$ 0.32 mm.}
\label{Fig.3} 
\end{figure}

 The I-N transition is of first order. When the transition proceeds via 
nucleation of bubbles, we observe long horizontal strings (as in ref.
\cite{lc2}), which are not suitable for our present analysis.  We selected 
those events where the transition seems to occur uniformly in a thin layer 
near the top of the droplet (possibly due to faster cooling from contact with 
air).  The depth of field of our microscope was about 20 microns. All the 
defects in the field of view were well focused suggesting that they 
formed in a thin layer, especially since typical inter-defect
separation was about 10 - 40 microns. (For us, the only thing relevant 
is that the layer be effectively two dimensional over distances of
order of typical inter-defect separation).  Also, the transition happened 
over the entire observation region roughly uniformly, suggesting that a 
process like spinodal decomposition may have been responsible for the 
transition. This resulted in a distribution of strength $\pm 1$ defects  
as shown in the photograph in Fig.3. Points
from which four dark brushes emanate correspond to defects of strength 
$\pm$ 1. Due to resolution limitation the crossings here do not appear as 
point like. It is practically impossible to use the technique of rotation 
of brushes to identify every winding in situations such as shown in Fig.3 
due to very small inter-defect separation (resulting from high defect
density), as well as due to rapid evolution of the defect distribution.
 
 We have developed a particular technique for determining individual 
windings of defects in situations like Fig.3 where one only needs to 
determine the winding of one of the defects by rotation in a cross 
polarizer setup. Windings of the rest of the defects can then be 
determined using topological arguments, see ref. \cite{lc3} for details
of this technique. Further, as explained in ref. \cite{lc3}, the anchoring 
of the director at the I-N interface \cite{angle} forces the director to 
lie on a cone, with the half angle equal to about 64$^0$. This forces the 
order parameter space there to become effectively a circle $S^1$, instead of 
being $RP^2$, with the order parameter being an angle between 0 and $2\pi$. 
Only defects allowed now are with integer windings which is consistent
with the fact that no strength 1/2 defects are seen in our experiment
\cite{lc3}. Therefore the prediction of the Kibble mechanism for the 
U(1) case, as described above, are valid for this case, with the 
picture that a domain structure near the I-N interface is responsible
for the formation of integer windings (see, ref. \cite{lc3} for detailed
discussion of these points).

 After identifying the windings of all defects (wherever possible) in a 
picture, we note down the positions (x-y coordinates) of each defect and 
anti-defect in the picture. We then determine the average defect/anti-defect 
density $\rho$. With the average inter-defect separation $r_{av}$ being 
$\rho^{-1/2}$, we convert all coordinate distances into scaled distances
by dividing with $r_{av}$. We then choose one defect as the origin,
making sure that the defect is at least 1 unit (with unit length being
$r_{av}$ now) away from the boundary of the picture.  Density of defects 
$\rho_d(r)$ as well as density of anti-defects $\rho_{\bar d}(r)$ is now 
calculated at each $r$ (at step size of $\Delta r$) 
by counting number of defects (anti-defects) within 
an annular region of thickness $\Delta r$ centered at $r$. Different values
of $\Delta r$ are used to to get least possible statistical fluctuations,
while at the same time, retaining the structure of the peak etc. Same
$\Delta r$ is used in numerical simulation for comparison with the
experimental data. We will present results for $\Delta r = 0.25$ (in the
units of $r_{av}$). The calculation is repeated by taking other defects 
as origins, and an average of densities (at a given value of $r$) is 
taken to determine final values of $\rho_d(r)$ and $\rho_{\bar d}(r)$. 
To increase the statistics, an average of the distributions is taken by 
combining the results of all pictures. Different pictures have very different
values of $r_{av}$, ranging from about 5 microns to about 50 microns.
However, when expressed in the units of $r_{av}$, the densities $\rho_d$
and $\rho_{\bar d}$ show similar behavior for all the pictures (at least
those ones which had significant statistics). In all, we have analyzed 
17 pictures, with total number of defects and anti-defects being 833. 

\begin{figure}[h]
\begin{center}
\leavevmode
\epsfysize=15truecm \vbox{\epsfbox{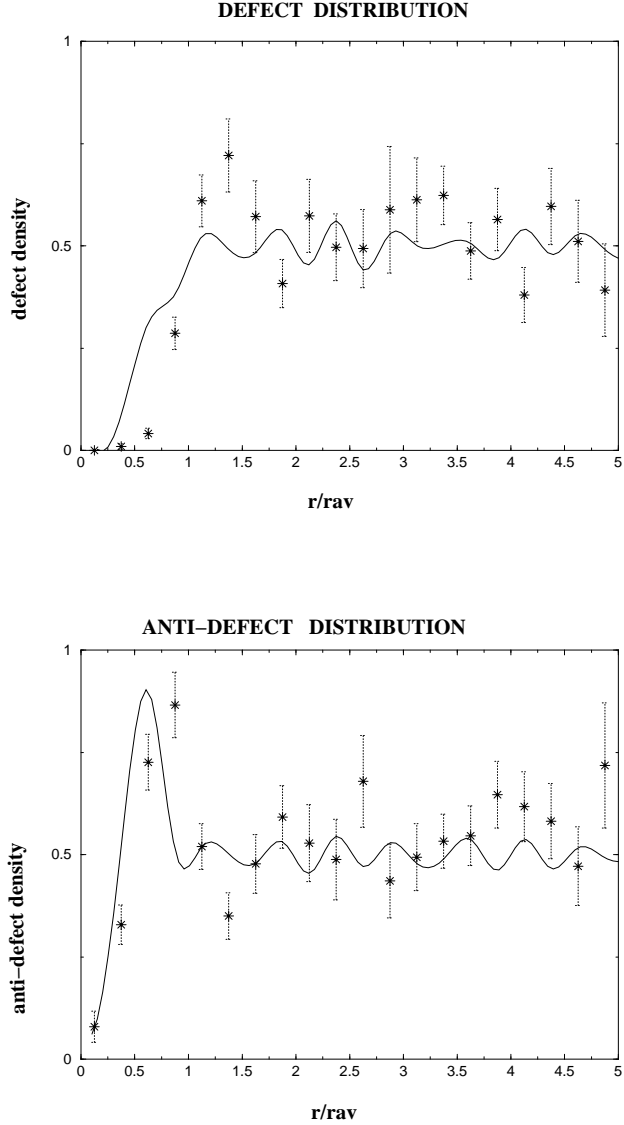}}
\end{center}
\caption{Solid plots show simulation results for defect
density $\rho_d(r)$ (top figure) and anti-defect density
$\rho_{\bar d}(r)$ (bottom figure). Stars show the experimental data.}
\label{Fig.4} 
\end{figure}

  Fig.4 shows the results. Solid plots shows the results of the simulations.
Antidefect distribution clearly shows a peak at $r \simeq 0.6$ as expected
from the Kibble mechanism. Defect density is suppressed in that region
at $r \simeq 0.6$, with the simulation results not too different from the 
expected suppression, i.e. $\rho_d \simeq \rho_{\bar d} - 0.4 \simeq
0.5$ (with $p \simeq 0.33$ and $\rho_{\bar d}(r = 0.57) \simeq 0.9$).
(As we mentioned above, accounting for difference between the corner
4 squares from the nearest 4 squares would have led to larger suppression
for $\rho_d(r = 0.57)$.) As expected, both densities
reach asymptotic values by $r \sim 1$. The regular oscillations in
the densities is due to the regular lattice structure of the simulations,
with defects and anti-defects remaining close to the centers of the domains.
If defects/anti-defects are assumed to be strictly at the centers of the
domains, then the oscillations in $\rho_{\bar d}(r)$ and $\rho_d(r)$ are
even more pronounced and regular. At the other extreme, if the positions of 
defects/anti-defects are taken to be uniform within respective domains, then
these oscillations completely disappear, except for the prominent peak in
$\rho_{\bar d}$ at $r \simeq 0.6$. Here we have taken an intermediate
case with the positions of defects/anti-defects weighted by a Gaussian 
centered at the middle of the elementary square (so that the peak height 
is similar to the one obtained from the experiment, as discussed above).

  Stars show the experimental values. The error bars have been calculated 
by taking the error in the count $n$ of defects or anti-defects within an 
annular strip (of the same thickness $\Delta r = 0.25$ used for the 
simulation) to be $\sqrt{n}$. The peak in the data points of anti-defect 
density is prominent, and so is the suppression in the defect density for 
$r < 1.0$. Data on defect density seems to be in a reasonably good
agreement with the simulation results, especially the amount of suppression
for $r < 1$.  The position of the anti-defect
peak from experimental data seems to be shifted by about 0.25 on right,
as compared to the peak from the simulation. With the statistics we have
at present, it is not possible to resolve whether this shift is genuine,
or whether it is due to statistical fluctuations. We have done following
checks to address this point. Instead of square elementary domains,
if we take triangular domains in the simulation (with due account of
geometrical factors in Eq.(2) etc.), we find the  simulation peak 
at about $r \simeq 0.4$, that is, slightly further shifted towards 
left compared to the experimental peak. This is consistent with the 
findings in ref.\cite{lc3} where it was found that experimental data favored
square elementary domains. Again, just as in the case in ref.\cite{lc3},
data in the present analysis also does not have enough statistics
to make definitive statements about this issue of preferred
shape of elementary domains. Even though, it is possible that the 
simulation peak may shift further to right for elementary domains with
larger number of sides (which increases the probability $p$),
it is clear from Eq.2 that the position of the 
peak will always remains at $r < 1$ (in units of $r_{av}$).

 With a smaller set of data we had seen that the shift between experimental
peak and the simulation peak was larger. With the inclusion of all
defects/anti-defects (which could be analyzed using our techniques),
the shift reduced, suggesting that it is possible that the shift may
reduce further if larger data is available. Even at present, the shift
between the two peaks is relatively small. In fact the shift is
about the same as the smallest separation between defects and/or anti-defects
which we have found in our experiments. 

 The observation of the peak in $\rho_{\bar d}(r)$, near $r \simeq 
0.8$, is roughly in accordance with the theoretical prediction. What is 
remarkable is that the data shows a prominent peak near the position 
where it is expected, and at the same time shows suppression in defect 
density by about the right amount, at the same point. At large $r$ there 
is not sufficient statistics to say whether the densities approach the 
asymptotic values at $r \simeq 1.1$, though the data is certainly 
consistent with this (in the sense that there are no other prominent 
peaks visible, and the fluctuations are randomly distributed about the 
asymptotic value). (We note here, that it is intriguing that similar structure 
of plots has been seen for the scaled radial distribution functions of 
islands formed on a single crystal substrate \cite{sv}. It will be 
interesting to explore of any possible connection between the two cases.)

 We conclude by stressing that our measurements of the density
correlation function of defects and anti-defects provide rigorous tests
of the theory of cosmic defect formation, as well as defect formation in
condensed matter systems. For a liquid crystal system, $r_{av}$ is about 10 
microns, while $r_{av}$ will be about $10^{-30}$ cm for cosmic defects (those 
formed at Grand Unified Theory transition). However, when expressed
in the scaled length $r$ (by dividing by $r_{av}$), one expects in both cases,
a peak in anti-defect density at $r \simeq 0.6$ and flattening out by
$r \simeq 1$ (for U(1) case and for 2 dimensional cross-sections of defect
networks). Similarly defect density is predicted to be suppressed by
a calculable factor  at $r \simeq 0.6$, again flattening out by $r \simeq
1$. Experimental data verifies both these predictions, though fluctuations
due to small statistics are large. It is clearly  desirable to
be able to carry out experimental analysis with significantly larger
data set in order to improve the error bars, so that more definitive
statements can be made about the comparison between theory and experiments,
(e.g., about the apparent shift in the peak position).
What is very encouraging and remarkable is that by appropriately
focusing on predictions of theory of cosmic defect formation which
acquire universal behavior  by suitable change of length scales,
one is able to rigorously test these theories in ordinary condensed
matter experiments. 

 We are very thankful to Soma Dey, Sanatan Digal, Soma Sanyal, Supratim 
Sengupta, and Shikha Varma, for useful discussions and comments. AMS would 
like to acknowledge the hospitality of the Physics Dept. Univ. of California,
Santa Barbara while the paper was being written. His work at UCSB
was supported by NSF Grant No. PHY-0098395.



\end{document}